\documentclass{article}%
\usepackage{amsfonts}
\usepackage{amsmath}
\usepackage{amssymb}
\usepackage{graphicx}%
\providecommand{\U}[1]{\protect\rule{.1in}{.1in}}

\begin{document}

\title{The Classical Limit of Entropic Quantum Dynamics\thanks{Presented at MaxEnt
2016, the 36th International Workshop on Bayesian Inference and Maximum
Entropy Methods in Science and Engineering (July 10-15, 2016, Ghent, Belgium).
}}
\author{Anthony Demme and Ariel Caticha\\{\small Department of Physics, University at Albany--SUNY, Albany, NY 12222,
USA}}
\date{}
\maketitle

\begin{abstract}
The framework of entropic dynamics (ED)\ allows one to derive quantum
mechanics as an application of entropic inference. In this work we derive the
classical limit of quantum mechanics in the context of ED. Our goal is to find
conditions so that the center of mass (CM) of a system of $N$ particles
behaves as a classical particle. What is of interest is that $\hbar$ remains
finite at all steps in the calculation and that the classical motion is
obtained as the result of a central limit theorem. More explicitly we show
that if the system is sufficiently large, and if the CM\ is initially
uncorrelated with other degrees of freedom, then the CM follows a smooth
trajectory and obeys the classical Hamilton-Jacobi with a vanishing quantum potential.

\end{abstract}

\section{Introduction}

Entropic Dynamics (ED) is a framework that allows the formulation of dynamical
theories as applications of entropic methods of inference.\footnote{The
principle of maximum entropy as a method for inference can be traced to the
pioneering work of E. T. Jaynes. For a pedagogical overview of Bayesian and
entropic inference and for further references see \cite{Caticha 2012}.} Its
main application so far has been the derivation of the Schr\"{o}dinger
equation \cite{Caticha 2010a}-\cite{Caticha 2015}. At this stage in its
development, the value of ED has been clarifying conceptual issues such as the
uncertainty relations, the quantum measurement problem, the connection to
Bohmian mechanics, etc. (see \cite{Caticha 2015} and references therein).
Indeed, given the experimental success of non-relativistic quantum mechanics
one should not expect any predictions that deviate from those of the standard
quantum theory --- at least not in the non-relativistic regime discussed here.
This situation will probably change once ED is extended to realms, such as
gravity, cosmology, or very high energies, where the status of quantum theory
is less well tested.

Whenever we reason on the basis of incomplete information it is inevitable
that the inference of most quantities will be afflicted by uncertainty. There
are however situations where for some very specially chosen variables one can,
despite the lack of information, achieve complete predictability. One
prominent example is thermodynamics where on the basis of just a few pieces of
information one can make precise predictions about the macroscopic behavior of
systems with huge numbers of particles. The catch, of course, is that the
answers to the vast majority of the questions one might conceivably ask about
the system (such as what is the position of a particular molecule) will remain
completely unanswered. However, for some very special variables ---
macrovariables such as temperature, pressure, heat capacity, etc. --- the
incomplete information happens to be sufficient to attain remarkable levels of
certainty. It is this phenomenon, which relies on the very careful choice of
special variables, that accounts for the emergence of determinism in theories
such as statistical mechanics or entropic dynamics that are intrinsically indeterministic.

In this paper we are concerned with studying a similar phenomenon, the
emergence of a deterministic classical mechanics from an intrinsically
indeterministic ED. The classical limit is usually taken by letting
$\hbar\rightarrow0$ \cite{Landau 1965}. This is mathematically correct, but
does not address the physical fact that any realistic macroscopic object is
comprised of a large number of particles whose motion is described by a
quantum mechanics in which $\hbar$ is not zero. We consider the entropic
dynamics of a system of $N$ quantum particles and show that for large $N$ the
center of mass (CM) behaves as a classical particle. The result is not in
itself surprising but some features of the derivation might however be of
interest. One is that $\hbar$ remains finite at all steps in the derivation
which means that a mesoscopic or macroscopic object will behave classically
while all its component particles remain explicitly quantum mechanical.

Two other interesting features are more directly related to the formalism of
ED. In ED particles follow Brownian trajectories. For large $N$ we find, as a
consequence of the central limit theorem, that the CM follows a smooth
trajectory. This condition is clearly necessary for a fully classical motion
but it is definitely not sufficient. (Particles in Bohmian mechanics, for
example, also follow smooth trajectories. \cite{Bohm Hiley 1993}\cite{Holland
1993})

In ED the motion is described through two coupled equations, one is a
Fokker-Planck equation for the probability density and the other is an
equation for the phase of the wave function. The latter is a Hamilton-Jacobi
modified by the addition of a quantum potential the form of which is suggested
by considerations of information geometry \cite{Caticha 2014b}. It is the
presence of this quantum potential that is ultimately responsible for quantum behavior.

We find that under rather general conditions the CM\ motion decouples, and for
a sufficiently large system the quantum potential for the CM\ motion vanishes.
Thus, for $N\rightarrow\infty$, the CM follows smooth trajectories described
by a classical Hamilton-Jacobi equation.\ In conclusion, the CM motion of a
system with a large number of particles follows a classical trajectory.

We begin with a brief overview of ED \cite{Caticha 2014b}. In section 3 we use
the central limit theorem to discuss the ED of the CM following the discussion
in \cite{Caticha 2012}\cite{Johnson 2011}. The Hamilton-Jacobi equation for
the CM motion, which is the new contribution of this paper, is derived in
section 4. We conclude in section 5 with some final remarks.

\section{Entropic Dynamics -- a brief review}

Our system consists of $N$ particles living in a flat Euclidean space
$\mathbf{X}$ with metric $\delta_{ab}$. The particles have definite positions
$x_{n}^{a}$. (The index $n$ $=1\ldots N$ denotes the particle and $a=1,2,3$
the spatial coordinate.) The position of the system in configuration space
$\mathbf{X}_{N}=\mathbf{X}\times\ldots\times\mathbf{X}$ --- what one might
call a microstate --- is denoted $x^{A}$ where $A=(n,a)$.

The goal is to predict the particle positions and their motion on the basis of
some limited information. The main piece of information is that particles
follow continuous trajectories which means that the motion can be analyzed as
a sequence of short steps. Thus, the first goal is to find the probability
$P(x^{\prime}|x)$ that the system takes an infinitesimally short step from
$x^{A}$ to $x^{\prime A}=x^{A}+\Delta x^{A}$. The tool to calculate this
probability is the method of maximum entropy and the relevant information is
introduced through constraints.

The information that particles take \emph{short} steps from $x_{n}^{a}$ to
$x_{n}^{\prime a}=x_{n}^{a}+\Delta x_{n}^{a}$ is expressed through $N$
independent constraints,
\begin{equation}
\langle\Delta x_{n}^{a}\Delta x_{n}^{b}\rangle\delta_{ab}=\kappa_{n}%
~,\qquad(n=1\ldots N)~.~\label{kappa n}%
\end{equation}
where we shall eventually take the limit $\kappa_{n}\rightarrow0$. But
particles do not move independently of each other. In order to introduce
correlations among them one imposes one additional constraint,
\begin{equation}
\langle\Delta x^{A}\rangle\partial_{A}\phi=\sum\limits_{n=1}^{N}\left\langle
\Delta x_{n}^{a}\right\rangle \frac{\partial\phi}{\partial x_{n}^{a}}%
=\kappa^{\prime}~,\label{kappa prime}%
\end{equation}
where $\phi$ is called the \emph{drift potential} and $\partial_{A}%
=\partial/\partial x^{A}=\partial/\partial x_{n}^{a}$. $\kappa^{\prime}$ is
another small but for now unspecified position-independent constant.
Eq.(\ref{kappa prime}) is a single constraint; it acts on the $3N$-dimensional
configuration space and is ultimately responsible for such quantum effects as
interference and entanglement. The physical nature of the $\phi$ potential
need not be discussed at this point --- it is sufficient to postulate its
existence and to note that it will eventually be transformed into the phase of
the wave function.\footnote{Since $\phi$ affects the motion of particles it
plays a role analogous to that of a pilot wave or an electromagnetic field.
Indeed, $\phi$ is \emph{as real as} the vector potential $A^{a}$ and the
intimate relation between the two manifests itself through a gauge symmetry
(see \cite{Caticha 2010a}\cite{Caticha 2015}). In \cite{Caticha 2010a} $\phi$
was interpreted as being itself of entropic origin and, in the context of
particles with spin, it is possible to interpret $\phi$ as an angular
variable. Clearly much remains to be clarified here.}

The result of maximizing entropy leads to
\begin{equation}
P(x^{\prime}|x)=\frac{1}{\zeta}\exp[-\sum_{n}(\frac{1}{2}\alpha_{n}\,\Delta
x_{n}^{a}\Delta x_{n}^{b}\delta_{ab}-\alpha^{\prime}\Delta x_{n}^{a}%
\frac{\partial\phi}{\partial x_{n}^{a}})]~, \label{Prob xp/x a}%
\end{equation}
where $\zeta$ is a normalization constant and $\alpha_{n}$ and $\alpha
^{\prime}$ are Lagrange multipliers. The specification of $\alpha^{\prime}$ is
relatively simple: one can rescale $\alpha^{\prime}\phi\rightarrow\phi$ which
amounts to choosing $\alpha^{\prime}=1$ without affecting the quantum dynamics
\cite{Bartolomeo Caticha 2016}. The specification of $\alpha_{n}$ is
considerably more involved. The limit of short steps ($\kappa_{n}\rightarrow
0$) is attained for $\alpha_{n}\rightarrow\infty$.

Once the probability $P(x^{\prime}|x)$ for an infinitesimal step is found one
proceeds by iteration to derive a Fokker-Planck equation for the probability
distribution $\rho(x,t)$. This requires the introduction of an
\textquotedblleft entropic\textquotedblright\ time $t$ as a book keeping
device to keep track of the accumulation of changes \cite{Caticha
2010a}-\cite{Caticha 2015}. The \textquotedblleft clock\textquotedblright%
\ that measures this entropic time is provided by the particle fluctuations.
These ideas are implemented by choosing
\begin{equation}
\alpha_{n}=\frac{m_{n}}{\eta\Delta t}~, \label{alpha n}%
\end{equation}
where the $m_{n}$'s are particle-specific constants that will be called
\textquotedblleft masses\textquotedblright\ and $\eta$ is a constant that
fixes the units of time relative to those of length and mass.

As discussed in \cite{Caticha 2014b}, up to an arbitrary scale factor $C$ the
geometry of the configuration space $\mathbf{X}_{N}$ is uniquely determined by
the information metric,\footnote{For an introduction to information geometry
with references to the literature see \emph{e.g.}, \cite{Caticha 2012}.}
\begin{equation}
m_{AB}=C\int dx^{\prime}\,P(x^{\prime}|x)\frac{\partial\log P(x^{\prime}%
|x)}{\partial x^{A}}\frac{\partial\log P(x^{\prime}|x)}{\partial x^{B}%
}~.\label{gamma C}%
\end{equation}
In the limit $\Delta t\rightarrow0$, with an appropriate choice of $C$, the
metric $m_{AB}$ is recognized as the \textquotedblleft mass\textquotedblright%
\ tensor while its inverse $m^{AB}$ is proportional to the \textquotedblleft
diffusion\textquotedblright\ tensor,
\begin{equation}
m_{AB}=m_{n}\delta_{AB}\quad\text{and}\quad m^{AB}=\frac{1}{m_{n}}\delta
^{AB}~.\label{mass tensor}%
\end{equation}

The choice of $\alpha_{n}$ in (\ref{alpha n}) leads to a simple dynamics:
$P(x^{\prime}|x)$ in eq.(\ref{Prob xp/x a}) is a Wiener process. A generic
displacement $\Delta x^{A}$ is expressed as an expected drift plus a
fluctuation,
\begin{equation}
\Delta x^{A}=b^{A}\Delta t+\Delta w^{A}~, \label{Delta x}%
\end{equation}
where $b^{A}(x)$ is the drift velocity,
\begin{equation}
\langle\Delta x^{A}\rangle=b^{A}\Delta t\quad\text{with}\quad b^{A}=\eta
m^{AB}\partial_{B}\phi~, \label{drift velocity}%
\end{equation}
and the f{}luctuations $\Delta w^{A}$ satisfy,
\begin{equation}
\langle\Delta w^{A}\rangle=0\quad\text{and}\quad\langle\Delta w^{A}\Delta
w^{B}\rangle=\frac{\eta}{m_{n}}\delta^{AB}\Delta t=\eta m^{AB}\Delta t~.
\label{fluc}%
\end{equation}

Having introduced a convenient notion of time through (\ref{alpha n}), the
accumulation of many changes leads to a Fokker-Planck equation for the
probability distribution $\rho(x,t)$, \cite{Caticha 2010a}\cite{Caticha
2014a}
\begin{equation}
\partial_{t}\rho=-\partial_{A}\left(  \rho v^{A}\right)  ~, \label{FP b}%
\end{equation}
where $v^{A}$ is the velocity of the probability flow in configuration space
or \emph{current velocity},
\begin{equation}
v^{A}=b^{A}+u^{A}\quad\text{and}\quad u^{A}=-\eta m^{AB}\partial_{B}\log
\rho^{1/2}~
\end{equation}
is the \emph{osmotic velocity}. Since both $b^{A}$ and $u^{A}$ are gradients,
the current velocity is a gradient too,%
\begin{equation}
v^{A}=m^{AB}\partial_{B}\Phi\quad\text{where}\quad\Phi=\eta\phi-\eta\log
\rho^{1/2}~. \label{curr}%
\end{equation}

The FP equation (\ref{FP b}) describes a standard diffusion. In order to
obtain a \textquotedblleft non-dissipative\textquotedblright\ dynamics
\cite{Nelson 1979} one must revise or update the constraint \ref{kappa prime}
after each step $\Delta t$. The result is that the drift potential $\phi$ (or
equivalently $\Phi$) becomes a dynamical degree of freedom instead of an
externally prescribed field. The required updating $\Phi\rightarrow\Phi
+\delta\Phi$ is such that a certain functional $\tilde{H}[\rho,\Phi]$ is
conserved,
\begin{equation}
\tilde{H}[\rho+\delta\rho,\Phi+\delta\Phi]=\tilde{H}[\rho,\Phi]~.
\end{equation}
The requirement that $\tilde{H}$ be conserved for arbitrary choices of $\rho$
and $\Phi$ implies that the coupled evolution of $\rho$ and $\Phi$ is given by
a conjugate pair of Hamilton's equations, \cite{Caticha 2014b}
\begin{equation}
\partial_{t}\rho=\frac{\delta\tilde{H}}{\delta\Phi}\quad\text{and}%
\quad\partial_{t}\Phi=-\frac{\delta\tilde{H}}{\delta\rho}~. \label{Hamilton}%
\end{equation}
The form of the \textquotedblleft ensemble\textquotedblright\ Hamiltonian
$\tilde{H}$ is chosen so that the first equation reproduces the FP equation
(\ref{FP b}). Then, the second equation in (\ref{Hamilton}) becomes a
Hamilton-Jacobi equation. A more complete specification of $\tilde{H}$ is
suggested by information geometry. The natural choice is
\begin{equation}
\tilde{H}[\rho,\Phi]=\int dx\,\left[  \frac{1}{2}\rho m^{AB}\partial_{A}%
\Phi\partial_{B}\Phi+\rho V+\xi m^{AB}\frac{1}{\rho}\partial_{A}\rho
\partial_{B}\rho\right]  ~, \label{Hamiltonian}%
\end{equation}
where the first term in the integrand is the \textquotedblleft
kinetic\textquotedblright\ term that reproduces (\ref{FP b}). The second term
includes the standard potential $V(x)$, and the third term, which is motivated
by information geometry, is called the \textquotedblleft
quantum\textquotedblright\ potential. The parameter $\xi=\hbar^{2}/8$\ defines
the value of what we call Planck's constant $\hbar$ \cite{Caticha 2014b}.

The formulation of ED is now complete. Its equivalence to quantum mechanics is
verified by combining $\rho$ and $\Phi$ into a single complex function,
\begin{equation}
\Psi=\rho^{1/2}\exp(i\Phi/\hbar)\quad\text{where}\quad\hbar=(8\xi)^{1/2}.
\label{psi k}%
\end{equation}
Then Hamilton's equations (\ref{Hamilton}) can be written as a single complex
linear Schr\"{o}dinger equation,%
\begin{equation}
i\hbar\partial_{t}\Psi=-\frac{\hbar^{2}}{2}m^{AB}\partial_{A}\partial_{B}%
\Psi+V\Psi~. \label{sch c}%
\end{equation}

\section{A Fokker-Planck equation for the Center of Mass}

As seen in eqs. (\ref{Delta x}) and (\ref{fluc}) the particle trajectories are
afflicted by fluctuations; ED is intrinsically indeterministic. To achieve a
classical limit it is necessary to suppress these fluctuations which can be
formally accomplished by making $\eta$ (or $\hbar$) sufficiently small or by
making $m$ sufficiently large. A more realistic approach \cite{Johnson 2011}
is to consider the motion of a very special macro-variable, the center of mass
of a large body composed of $N$ particles. The goal is to show that while the
positions and motions of all the component particles remain uncertain, the CM
motion becomes fully predictable in the large $N$ limit.

From (\ref{Prob xp/x a}) with (\ref{alpha n}) the transition probability for a
short step is
\begin{equation}
P(x^{\prime}|x)=\frac{1}{Z_{N}}\exp\left[  -%
{\textstyle\sum\limits_{n=1}^{N}}
\frac{m_{n}}{2\eta\Delta t}\delta_{ab}\left(  \Delta x_{n}^{a}-\Delta\bar
{x}_{n}^{a}\right)  (\Delta x_{n}^{b}-\Delta\bar{x}_{n}^{b})\right]  ~.
\label{Prob xp/x b}%
\end{equation}
The CM\ coordinates are
\begin{equation}
X^{a}=\frac{1}{M}%
{\textstyle\sum\limits_{n=1}^{N}}
m_{n}x^{a}\quad\text{where}\quad M=N\bar{m}=%
{\textstyle\sum\limits_{n=1}^{N}}
m_{n}~. \label{CM coords}%
\end{equation}
\noindent\noindent The probability that the center of mass moves from $X$ to
$X^{\prime}=X+\Delta X$ is found from (\ref{Prob xp/x b}),%
\begin{equation}
P(X^{\prime}|X)=\int d^{3N}x^{\prime}\,P(x^{\prime}|x)\delta\left(  \Delta
X-\frac{1}{M}%
{\textstyle\sum\limits_{n}}
m_{n}\Delta x\right)  ~.
\end{equation}
The evaluation of $P(X^{\prime}|X)$ is a straightforward consequence of the
central limit theorem (see \emph{e.g. }\cite{Caticha 2012}). The result is
\begin{equation}
P(X^{\prime}|X)\propto\exp\left[  -\frac{M}{2\eta\Delta t}\delta_{ab}\left(
\Delta X^{a}-\Delta\bar{X}^{a}\right)  \left(  \Delta X^{b}-\Delta\bar{X}%
^{b}\right)  \right]  ~.
\end{equation}
Therefore, a short step of the CM\ is given by
\begin{equation}
\Delta X^{a}=\Delta\bar{X}^{a}+\Delta W^{a}~,
\end{equation}
where, using eq.(\ref{Delta x}) and (\ref{drift velocity}), the expected step
$\Delta\bar{X}^{a}$ is
\begin{equation}
\Delta\bar{X}^{a}=\frac{1}{M}%
{\textstyle\sum\limits_{n=1}^{N}}
m_{n}\Delta\bar{x}^{a}=\frac{\eta\Delta t}{N\bar{m}}%
{\textstyle\sum\limits_{n=1}^{N}}
\frac{\partial\phi}{\partial x_{n}^{a}}~, \label{CM drift}%
\end{equation}
and the fluctuations $\Delta W^{a}$ such that
\begin{equation}
\left\langle \Delta W^{a}\Delta W^{b}\right\rangle =\frac{\eta}{N\bar{m}%
}\Delta t\,\delta^{ab}~. \label{CM fluc}%
\end{equation}
Thus, we see that the expected drift $\Delta\bar{X}^{a}$ is of order $N^{0}$
(because the $N$ terms in the sum offset the $N$ in the denominator) whereas
the fluctuations are of order $N^{-1/2}$. For large $N$ the fluctuations
become negligible and the CM follows a smooth trajectory.

Equation (\ref{CM drift}) can be simplified considerably by introducing the
internal coordinates $\hat{x}_{n}^{a}$ relative to the CM,
\begin{equation}
x_{n}^{a}=X^{a}+\hat{x}_{n}^{a}\quad\text{so that}\quad%
{\textstyle\sum\limits_{n=1}^{N}}
m_{n}\hat{x}_{n}^{a}=0~.\label{relative coords}%
\end{equation}
Differentiate
\begin{equation}
\phi(x_{1}^{a}\ldots x_{N}^{a})=\phi(X^{a}+\hat{x}_{1}^{a},\ldots,X^{a}%
+\hat{x}_{N}^{a})
\end{equation}
keeping $\hat{x}$ constant to get
\begin{equation}
\frac{\partial\phi}{\partial X^{a}}=%
{\textstyle\sum\limits_{n=1}^{N}}
\frac{\partial\phi}{\partial x_{n}^{a}}\frac{\partial x_{n}^{a}}{\partial
X^{a}}=%
{\textstyle\sum\limits_{n=1}^{N}}
\frac{\partial\phi}{\partial x_{n}^{a}}~.\label{momentum a}%
\end{equation}
Then eqs. (\ref{CM drift}) and (\ref{CM fluc}) become%
\begin{equation}
\Delta\bar{X}^{a}=\frac{\eta}{M}\frac{\partial\phi}{\partial X^{a}}\Delta
t=B^{a}\Delta t\quad\text{and}\quad\left\langle \Delta W^{a}\Delta
W^{b}\right\rangle =\frac{\eta}{M}\Delta t\,\delta^{ab}%
~.\label{CM drift and fluc}%
\end{equation}

Now, in exactly the same way \cite{Caticha 2012} that the FP eq.(\ref{FP b})
can be derived from eqs.(\ref{drift velocity}) and (\ref{fluc}), we can use
(\ref{CM drift and fluc}) to derive a FP\ equation for the probability
$\rho_{CM}(X,t)$ of the CM. The result is
\begin{equation}
\partial_{t}\rho_{CM}=-\partial_{a}\left(  \rho_{CM}V^{a}\right)  ~,
\label{CM FPeq}%
\end{equation}
where $V^{a}$ is the CM \emph{current }velocity,
\begin{equation}
V^{a}=B^{a}+U^{a}\quad\text{and}\quad U^{a}=-\frac{\eta}{M}\partial_{B}%
\log\rho^{1/2}~
\end{equation}
is the CM \emph{osmotic }velocity. Notice that for large $N$ or $M\ $the
osmotic contribution vanishes, there is no diffusion and the probability
distribution $\rho_{CM}$ flows along the drift velocity. This is exactly what
one expects in classical mechanics. For large $N$ (\ref{CM drift and fluc})
shows that the CM\ trajectory follows the gradient of the scalar function,
\begin{equation}
M\frac{dX^{a}}{dt}=M\frac{d\bar{X}^{a}}{dt}=\eta\frac{\partial\phi}{\partial
X^{a}}~.
\end{equation}
which is the classical equation of motion of a particle of mass $M$ in the
Hamilton-Jacobi formalism.

As mentioned earlier this condition is necessary to obtain the classical limit
but it is not sufficient. To complete the argument we must show that the
scalar function $\eta\phi$ obeys the classical HJ equation.

\section{The Hamilton-Jacobi equation for the Center of Mass}

We start from eqs.(\ref{Hamilton}) and (\ref{Hamiltonian}) to get the HJ
equation for the $N$ particles,%
\begin{equation}
-\partial_{t}\Phi=\frac{\delta\tilde{H}}{\delta\rho}=\frac{1}{2}m^{AB}%
\partial_{A}\Phi\partial_{B}\Phi+V+V^{Q}~. \label{HJb}%
\end{equation}
The quantum potential $V_{Q}$ is given by
\begin{equation}
V^{Q}=\frac{\delta}{\delta\rho}\xi m^{AB}I_{AB}[\rho]\quad\text{where}\quad
I_{AB}[\rho]=\int dx\,\frac{1}{\rho}\partial_{A}\rho\partial_{B}\rho~,
\label{q pot}%
\end{equation}
which leads to
\begin{equation}
V^{Q}=-4\xi m^{AB}\frac{\partial_{A}\partial_{B}\rho^{1/2}}{\rho^{1/2}}~.
\end{equation}
Next we change from $x_{n}^{a}$ ($n=1\ldots N$) to the CM coordinates
$(X^{a},\hat{x}_{\ell}^{a})$ ($\ell=1\ldots N-1$).

\paragraph*{Changing to CM coordinates}

We need%
\begin{equation}
\frac{\partial\Phi}{\partial x_{n}^{a}}=\frac{\partial\Phi}{\partial X^{b}%
}\frac{\partial X^{b}}{\partial x_{n}^{a}}+%
{\textstyle\sum\limits_{\ell}^{N-1}}
\frac{\partial\Phi}{\partial\hat{x}_{\ell}^{b}}\frac{\partial\hat{x}_{\ell
}^{b}}{\partial x_{n}^{a}}~,
\end{equation}
Use (\ref{CM coords}) and (\ref{relative coords}), to write%
\begin{equation}
\frac{\partial X^{b}}{\partial x_{n}^{a}}=\frac{m_{n}}{M}\delta_{a}^{b}%
\end{equation}
and
\begin{equation}
\frac{\partial\hat{x}_{\ell}^{b}}{\partial x_{n}^{a}}=\delta_{a}^{b}\left(
\delta_{\ell n}-\frac{m_{n}}{M}\right)  \quad\text{for}\quad\ell=1\ldots N~.
\end{equation}
Then
\begin{equation}
\frac{\partial\Phi}{\partial x_{n}^{a}}=\frac{m_{n}}{M}\frac{\partial\Phi
}{\partial X^{a}}+%
{\textstyle\sum\limits_{\ell}^{N-1}}
\frac{\partial\Phi}{\partial\hat{x}_{\ell}^{a}}\left(  \delta_{\ell n}%
-\frac{m_{n}}{M}\right)  ~. \label{partial x a}%
\end{equation}
Summing over $n$ gives,
\begin{equation}%
{\textstyle\sum\limits_{n=1}^{N}}
\frac{\partial\Phi}{\partial x_{n}^{a}}=%
{\textstyle\sum\limits_{n=1}^{N}}
\frac{m_{n}}{M}\frac{\partial\Phi}{\partial X^{a}}+%
{\textstyle\sum\limits_{\ell}^{N-1}}
\frac{\partial\Phi}{\partial\hat{x}_{\ell}^{a}}%
{\textstyle\sum\limits_{n=1}^{N}}
\left(  \delta_{\ell n}-\frac{m_{n}}{M}\right)  =\frac{\partial\Phi}{\partial
X^{a}}~,
\end{equation}
which is the analogue of (\ref{momentum a}),
\begin{equation}
\frac{\partial\Phi}{\partial X^{a}}=%
{\textstyle\sum\limits_{n=1}^{N}}
\frac{\partial\Phi}{\partial x_{n}^{a}}~. \label{momentum b}%
\end{equation}
In the HJ\ formalism this equation has a straightforward interpretation: it
expresses the CM momentum in terms of the momentum of the component particles.

\paragraph*{The kinetic energy}

Using (\ref{partial x a}) in the kinetic term in (\ref{HJb}),
\begin{align}
K  &  =%
{\textstyle\sum\limits_{n}^{N}}
\frac{\delta_{ab}}{m_{n}}\frac{\partial\Phi}{\partial x_{n}^{a}}\frac
{\partial\Phi}{\partial x_{n}^{b}}=\frac{1}{2M}\left(  \frac{\partial\Phi
}{\partial X^{a}}\right)  ^{2}+\nonumber\\
&  +%
{\textstyle\sum\limits_{\ell}^{N-1}}
\frac{1}{2m_{\ell}}\left(  \frac{\partial\Phi}{\partial\hat{x}_{\ell}^{a}%
}\right)  ^{2}-\frac{1}{2M}\left(
{\textstyle\sum\limits_{\ell}^{N-1}}
\frac{\partial\Phi}{\partial\hat{x}_{\ell}^{a}}\right)  ^{2}~,
\label{kinetic term}%
\end{align}
shows that the contributions of the CM and the internal coordinates decouple.

\paragraph*{The potentials}

It remains to study whether this decoupling survives the effects of the
potential $V$ and quantum potential $V_{Q}$ terms in (\ref{HJb}). The
situation with $V$ is straightforward: there exist broad families of
potentials of practical interest for which the CM and the internal coordinates
decouple. For example,
\begin{equation}
V(x)=V_{\text{ext}}(X)+\sum_{n,\ell}V_{\text{int}}(\hat{x}_{n}-\hat{x}_{\ell
})~, \label{potential}%
\end{equation}
where $V_{\text{ext}}$ and $V_{\text{int}}$ represent an external potential
and the interparticle interactions respectively.

The situation with the quantum potential (\ref{q pot}) is subtler. A generic
probability distribution
\begin{equation}
\rho(x)=\rho(X)\rho(\hat{x}|X)
\end{equation}
leads to quantum correlations between the CM and the internal coordinates. In
order for the CM\ motion to decouple it is necessary to invoke some
decoherence mechanism that leads to initial conditions where $X$ and $\hat{x}$
are independent,
\begin{equation}
\rho(x)=\rho_{CM}(X)\hat{\rho}(\hat{x})\ .\label{uncorrelated}%
\end{equation}
One can then see that if $X$ and $\hat{x}$ are initially uncorrelated then the
dynamics is such that they remain uncorrelated. To prove this we show that the
relevant term in (\ref{q pot}) decouples,
\begin{align}
m^{AB}I_{AB}[\rho] &  =%
{\textstyle\sum\limits_{n=1}^{N}}
\frac{1}{m_{n}}\int dx\,\rho~\left(  \frac{\partial\log\rho}{\partial
x_{n}^{a}}\right)  ^{2}\nonumber\\
&  =%
{\textstyle\sum\limits_{n=1}^{N}}
\frac{1}{m_{n}}\int dXd\hat{x}\,\rho_{CM}\hat{\rho}\left(  \frac{m_{n}}%
{M}\frac{\partial\log\rho_{CM}}{\partial X^{a}}+\frac{\partial\log\hat{\rho}%
}{\partial\hat{x}_{n}^{a}}\right)  ^{2}\nonumber\\
&  =\frac{1}{M}\int dX\,\rho_{CM}\left(  \frac{\partial\log\rho_{CM}}{\partial
X^{a}}\right)  ^{2}+%
{\textstyle\sum\limits_{n=1}^{N}}
\frac{1}{m_{n}}\int d\hat{x}\,\hat{\rho}\left(  \frac{\partial\log\hat{\rho}%
}{\partial\hat{x}_{n}^{a}}\right)  ^{2}.\label{q pot b}%
\end{align}
Therefore, the quantum potential becomes,
\begin{equation}
V^{Q}(x)=V_{CM}^{Q}(X)+\hat{V}_{CM}^{Q}(\hat{x})
\end{equation}
where
\begin{equation}
V_{CM}^{Q}(X)=-\frac{4\xi}{M\rho_{CM}^{1/2}}\delta^{ab}\frac{\partial^{2}%
\rho_{CM}^{1/2}}{\partial X^{a}\partial X^{b}}=-\frac{4\xi}{M\rho_{CM}^{1/2}%
}\nabla^{2}\rho_{CM}^{1/2}~,\label{q pot CM}%
\end{equation}
and
\begin{equation}
\hat{V}^{Q}(\hat{x})=-4\xi%
{\textstyle\sum\limits_{n=1}^{N}}
\frac{1}{m_{n}}\frac{1}{\hat{\rho}^{1/2}}\delta^{ab}\frac{\partial^{2}%
\hat{\rho}^{1/2}}{\partial\hat{x}_{n}^{a}\partial\hat{x}_{n}^{b}}~.
\end{equation}

Collecting these results, (\ref{kinetic term}), (\ref{potential}), and
(\ref{q pot b}) into (\ref{HJb}), the HJ becomes
\begin{align}
-\partial_{t}\Phi &  =\frac{1}{2M}\left(  \frac{\partial\Phi}{\partial X^{a}%
}\right)  ^{2}+V_{\text{ext}}(X)+V_{CM}^{Q}(X)\nonumber\\
&  +%
{\textstyle\sum\limits_{\ell}^{N-1}}
\frac{1}{2m_{\ell}}\left(  \frac{\partial\Phi}{\partial\hat{x}_{\ell}^{a}%
}\right)  ^{2}-\frac{1}{2M}\left(
{\textstyle\sum\limits_{\ell}^{N-1}}
\frac{\partial\Phi}{\partial\hat{x}_{\ell}^{a}}\right)  ^{2}+%
{\textstyle\sum\limits_{n,\ell}^{N}}
V_{\text{int}}(\hat{x}_{n}-\hat{x}_{\ell})+\hat{V}^{Q}(\hat{x})~.
\end{align}
Thus, under the conditions (\ref{potential}) and (\ref{uncorrelated}) the
HJ\ equation can be solved by separation of variables,
\begin{equation}
\Phi(x)=\Phi_{CM}(X)+\hat{\Phi}(\hat{x})~.
\end{equation}
Direct substitution leads to the HJ equation for the CM,
\begin{equation}
-\partial_{t}\Phi_{CM}=\frac{1}{2M}\left(  \frac{\partial\Phi_{CM}}{\partial
X^{a}}\right)  ^{2}+V_{\text{ext}}(X)+V_{CM}^{Q}(X)~. \label{CM HJ}%
\end{equation}

Inspection of the FP equation (\ref{CM FPeq}) and the HJ equation
(\ref{CM HJ}) shows that the CM\ motion is described by a standard
Schr\"{o}dinger equation. If the system is sufficiently large, $M\rightarrow
\infty$, we see from (\ref{q pot CM}) that $V_{CM}^{Q}$ is suppressed. In
conjunction with (\ref{CM drift and fluc}) which implies smooth trajectories
we see that the CM\ motion is described by classical mechanics. This concludes
our derivation.

\section{Final remarks}

We have derived the classical limit of entropic quantum dynamics. The
derivation hinges on two features that are of an intrinsically probabilistic
or inferential nature: one is the central limit theorem, the other is the
specific form of the quantum potential as dictated by information geometry.

The ED framework appears to be ideally suited for the study the transition
between classical and quantum mechanics. In future work we intend to study the
effect of large but still finite values of $N$. The diffraction of
macro-molecules, for example, might offer an interesting test case where $N$
might be small enough that the quantum potential is not totally negligible,
but large enough that a discussion in terms of a smooth CM trajectory might be
useful \cite{Arndt et al 2005}. In addition, being an inference theory,
entropic dynamics provides the natural framework to account for loss of
information through decoherence \cite{Zurek 2003}\cite{Giulini et al 2003}.
The results of this paper provide a first step in that direction.

\paragraph*{Acknowledgments}

We would like to thank M. Abedi, C. Cafaro, N. Caticha, S. DiFranzo, A.
Giffin, S. Ipek, D.T. Johnson, K. Knuth, S. Nawaz, M. Reginatto, C. Rodriguez,
and K. Vanslette, for many discussions on entropy, inference and quantum mechanics.

\end{document}